# Treillis de concepts et ontologies pour l'interrogation d'un annuaire de sources de données biologiques (BioRegistry)

N. Messai, M-D. Devignes, A. Napoli, M. Smaïl-Tabbone

*UMR 7503 LORIA, BP 239, 54506 Vandœuvre Lès Nancy, FRANCE.*
*{messai,devignes,napoli,malika}@loria.fr ; http://www.loria.fr/equipes/orpailleur*

RÉSUMÉ. *Les sources de données biologiques disponibles sur le web sont multiples et hétérogènes. L'utilisation optimale de ces ressources nécessite aujourd'hui de la part des utilisateurs des compétences à la fois en informatique et en biologie, du fait du manque de documentation et des difficultés d'interaction avec les sources de données. De fait, les contenus de ces ressources restent souvent sous-exploités. Nous présentons ici une approche basée sur l'analyse de concepts formels, pour organiser et rechercher des sources de données biologiques pertinentes pour une requête donnée. Le travail consiste à construire un treillis de concepts à partir des méta-données associées aux sources. Le concept construit à partir d'une requête donnée est alors intégré au treillis. La réponse à la requête est ensuite fournie par l'extraction des sources de données appartenant aux extensions des concepts subsumant le concept requête dans le treillis. Les sources ainsi retournées peuvent être triées selon l'ordre de spécificité des concepts dans le treillis. Une procédure de raffinement de requête, basée sur des ontologies du domaine, permet d'améliorer le rappel par généralisation ou par spécialisation*

ABSTRACT. *Bioinformatic data sources available on the web are multiple and heterogenous. The lack of documentation and the difficulty of interaction with these data sources require users competence in both informatics and biological fields for an optimal use of sources contents that remain rather under exploited. In this paper we present an approach based on formal concept analysis to classify and search relevant bioinformatic data sources for a given query. It consists in building the concept lattice from the binary relation between bioinformatic data sources and their associated metadata. The concept built from a given query is then merged into the concept lattice. The result is given by the extraction of the set of sources belonging to the extents of the query concept subsumers in the resulting concept lattice. The sources ranking is given by the concept specificity order in the concept lattice. An improvement of the approach consists in automatic query refinement thanks to domain ontologies. Two forms of refinement are possible by generalisation and by specialisation.*

MOTS-CLÉS : *méta-données, bioinformatique, treillis de concepts, ontologies, sources de données.*

# 1. Introduction

L'un des grands défis de la bioinformatique aujourd'hui est de permettre aux biologistes d'accéder efficacement aux données gisant dans les centaines de sources de données réparties à travers le monde. Le grand nombre de sources, leur hétérogénéité et la complexité des objets biologiques auxquels elles font référence rendent souvent difficile la mise en relation d'une requête avec la source appropriée. Souvent la requête elle-même doit être décomposée en requêtes élémentaires susceptibles d'être adressées à des sources distinctes. Des approches variées ont été adoptées pour unifier l'accès aux diverses sources de données face à une requête donnée. Des systèmes ont été produits, à partir d'entrepôts de données (tels que GUS (Davison et al., 2000)), de fédération de bases de données (ex : SEMEDA (Köhler et al., 2003)), ou de médiateurs (ex : TAMBIS (Goble et al., 2001)). Certains de ces systèmes comme TAMBIS ou SEMEDA sont capables d'interpréter de façon sémantique la requête. Cependant la plupart des systèmes disponibles ne prend en compte qu'un petit nombre de sources de données et ne peut donc satisfaire un large éventail de requêtes utilisateurs. Le travail présenté ici vise à modéliser les connaissances disponibles sur les sources de données biologiques afin de proposer aux utilisateurs les sources de données les mieux adaptées à leur requête. Le problème ici n'est pas l'interrogation des sources elles-mêmes mais plutôt l'identification et le choix parmi toutes les sources de données disponibles de celles qui sont les plus appropriées par rapport à un besoin donné. Dans cet article, nous proposons de considérer ce problème comme un cas particulier de Recherche d'Information (IR pour *Information Retrieval*) dans lequel des sources de données plutôt que des documents font l'objet de la recherche et où l'indexation utilise les méta-données associées aux sources de données plutôt que les termes extraits des documents. L'analyse de concepts formels (de l'anglais *Formal Concept Analysis* abrégé en FCA) est utilisée ici pour faciliter la recherche grâce à une classification flexible et dynamique des sources existantes. De plus les ontologies de domaines sont prises en compte au moment de l'indexation des sources, ce qui permet de traiter les requêtes de façon sémantique pour améliorer la recherche.

Nous commencerons par décrire et discuter brièvement en section 2 des travaux apparentés, impliquant FCA et IR, ainsi que l'usage d'ontologies, dans des problèmes similaires. La section 3 présentera le projet BioRegistry comme un annuaire structuré répertoriant les méta-données associées aux sources de données biologiques. La formalisation de notre problème grâce à la FCA sera détaillée en section 4 et les aspects relatifs à la recherche d'information seront développés en section 5, avec en particulier une technique originale de raffinement de requête. Une conclusion et quelques perspectives de ce travail seront présentées en section 6.

## 2. Travaux voisins

### 2.1. *Utilisation de treillis de concepts pour la recherche d'information*

Les treillis de concepts ont été appliqués dans la recherche d'information (IR) dès l'apparition de l'analyse de concepts formels (Wille, 1982). En effet une analogie évidente existe entre les tableaux objet-attribut et document-terme. La recherche d'information a par la suite été explicitement mentionnée comme étant l'une des application possibles des treillis de concepts (Godin et al, 1995). Les concepts formels sont vus comme des classes de documents pertinents pour une requête donnée. La relation de subsumption (la relation d'ordre partiel entre les concepts du treillis) entre les concepts permet le passage d'un concept (ou d'une requête) à un autre plus général ou plus spécifique. Une approche pour la recherche d'information en utilisant les treillis de concepts a été proposé dans (Carpineto et al., 2000). Dans les deux propositions (Godin et al, 1995) et (Carpineto et al., 2000), la recherche d'information utilisant les treillis de concepts atteint des performances qui dépassent nettement celles de la recherche booléenne classique. Une limite de l'utilisation des treillis de concepts pour la recherche d'information est la complexité du treillis (nombre de concepts) pour les grands contextes. Cependant dans les applications réelles la complexité théorique maximale n'est pas atteinte (Carpineto et al., 2000). De plus des solutions visant à contrôler la taille des treillis correspondant aux grands contextes ont été proposées dans (Pernelle et al., 2002) et (Stumme et al., 2001).

### 2.2. *Amélioration des performances en utilisant les ontologies de domaine*

Le raffinement de requête est un mécanisme visant à améliorer les performances de la recherche d'information en ajoutant à la requête utilisateur de nouveaux termes liés à ceux initialement présents dans la requête (Carmel et al., 2002). La combinaison ontologies et FCA dans le but d'améliorer les performances de la recherche d'information a fait l'objet des propositions (Carpineto et al., 2000, Priss, 2000, Safar et al., 2004). Dans les deux premières propositions, un thésaurus est utilisé pour améliorer les performances du processus de recherche d'information en enrichissant l'indexation dans le treillis. Dans la troisième proposition, les ontologies de domaine sont utilisées pour raffiner la construction du treillis selon les préférences des utilisateurs et ceci permet d'éviter la construction du treillis complet. Les deux approches modifient directement le treillis soit en ajoutant des termes d'indexation soit en ne considérant qu'une partie du treillis.

Dans notre travail, les ontologies de domaines sont prises en compte très tôt dans le processus de recherche d'information (dès la construction de BioRegistry). Ceci nous amène à proposer une méthode consistant à modifier la requête au lieu du treillis pour améliorer le processus de recherche d'information.

### 2.3. *Algorithmes de construction incrémentale de treillis de concepts*

Le problème de calcul des concepts d'un treillis de concepts à partir d'un contexte formel a fait l'objet de plusieurs travaux de recherche. Une comparaison détaillée des performances des algorithmes proposés pour la génération des treillis et de leurs diagrammes de Hasse correspondants est présentée dans (Kuznetsov et al., 2002). Parmi les algorithmes proposés, quelques-uns ont la spécificité d'effectuer une construction incrémentale des treillis de concepts à partir de contextes formels (Godin et al., 1995, Carpineto et al., 1996). Cet aspect est particulièrement intéressant pour l'application des treillis de concepts à la recherche d'information en général et à notre problème de recherche de sources de données biologiques en particulier. En effet, les requêtes utilisateurs peuvent être insérées dans le treillis représentant la collection de documents (ou de sources de données). Suite à cette insertion il est possible de déterminer les documents les plus pertinents répondants aux critères exprimés par l'utilisateur dans sa requête. La construction incrémentale des treillis de concepts permet aussi l'ajout de nouveaux concepts ce qui rend possible la prise en compte de l'apparition de nouvelles sources de données biologiques sur le web. Dans notre cas, cet ajout est essentiel pour maintenir à jour l'annuaire BioRegistry par rapport au contenu du web.

## 3. Le projet BioRegistry

### 3.1. *Les sources de données biologiques*

Des centaines de sources de données biologiques sont connues de nos jours (Galperin, 2005). De nombreux efforts ont été consacrés jusqu'à présent aux problèmes posés par l'unification de l'accès à ces sources, le traitement des requêtes en vue de leur distribution sur les sources pertinentes, l'intégration des réponses, etc. Ces tâches nécessitent de concevoir des scénarios (workflow) appropriés et de disposer d'une interopérabilité transparente entre les ressources. Des systèmes intégrés ont été développés à partir d'architectures d'entrepôts de données ou de systèmes de médiation. Les solutions actuelles sont envisagées aussi dans le contexte du web sémantique avec en particulier la composition des services web (Oinn et al., 2004). Pour que toutes ces solutions deviennent réellement efficaces, il semble souhaitable que l'ensemble des connaissances disponibles sur les sources de données puisse être exploité. Par exemple, une requête apparemment simple comme: « *Quels sont les gènes du chromosome X humain qui sont exprimés préférentiellement dans le cerveau ?* » traite de données dites *de cartographie* et *d'expression*, qui peuvent à un instant donné être contenues dans une source unique ou dans des sources distinctes. Vraisemblablement, il sera possible de trouver plusieurs sources pertinentes pour chaque type de données. L'utilisateur pourra aussi choisir l'une de ces sources à cause de critères de qualité particuliers tels que la révision manuelle des données ou la fréquence de mise à jour, ou encore en raison de contraintes d'accès aux données.

Le catalogue de sources de données biologiques le plus documenté aujourd'hui est certainement DBCAT (Discala et al., 2000). Cet annuaire au format de fichier plat contient un ensemble de méta-données relativement restreint sur plus de 400 sources de données. Les possibilités d'interrogation sont cependant limitées par le fait que la plupart des champs sont de domaine ouvert (texte libre). D'autres annuaires sont développés aujourd'hui pour les services web en bioinformatique comme par exemple dans les projets MyGrid et BioMoby (Lord et al., 2004, Wroe et al., 2003). La proportion d'information biologique accessible par les services web est encore trop limitée pour répondre aux besoins des utilisateurs. Cependant cette situation pourrait changer et le besoin de modéliser et d'organiser les connaissances concernant les services web en bioinformatique pourrait devenir aussi pressant qu'il l'est aujourd'hui pour les sources de données biologiques. Afin de disposer d'un environnement adéquat pour tester nos propositions concernant la classification et la recherche de sources de données biologiques, nous avons décidé de construire notre propre annuaire, baptisé BioRegistry, dans lequel les méta-données associées aux sources de données biologiques pourront être organisées de façon dynamique, flexible et structurée.

### 3.2. *Le modèle de méta-données de BioRegistry*

En 1995, un comité international d'experts a proposé un modèle standard de description de méta-données relatives aux ressources du web : le *Dublin Core Metadata Initiative* ou DCMI (Dekkers et al., 2003). Ce standard est composé d'une quinzaine de champs ou éléments dont les principaux sont : *title*, *creator*, *subject*, *description*, *publisher*, *type*, *language*, *rights*, *coverage/spatial*, *verage/temporal*, *identifier*, *relation*, *format*, *date/created*, *date/modified*.

Le modèle des méta-données du DCMI étant très général, son utilisation pour décrire des ressources d'un domaine particulier nécessite souvent des raffinements voire des extensions. Ainsi le FGDC[1] est le comité chargé de standardiser les méta-données sur les données géospatiales digitales. La richesse et la spécificité des sources de données biologiques nous ont également conduits à la proposition d'un modèle hiérarchique pour l'organisation des méta-données à attacher à ces ressources. Le modèle de BioRegistry comporte quatre catégories de méta-données :

– Identification de la source : nous retrouvons ici plusieurs champs du DCMI (tels *title*, *publisher*, *date/modified*, *coverage/temporal*) auxquels s'ajoutent d'autres champs comme la fréquence de mise à jour.

---

1. http://www.fgdc.gov/fgdc/fgdc.html

– Thèmes couverts par la source : nous retrouvons ici le champ *subject* du DCMI mais aussi un champ relatif aux organismes couverts par une source (ex. la source nommée *Mouse Genome DB* contient des données sur l'organisme *Mouse*).

– Qualité de la source de données : cette catégorie, absente du DCMI, est cruciale pour documenter la qualité d'une source biologique par rapport au mode de validation de ses entrées, la compatibilité par rapport aux standards, la couverture (nombre de *gènes*, nombre de *contigs*...) et l'existence de références croisées avec d'autres sources de données.

– Disponibilité de la source : cette catégorie regroupe les champs concernant les adresses des différents sites donnant accès à une source de donnée, ainsi que les contraintes d'accès pour les mondes académique et industriel (gratuité, authentification…).

Nous avons suivi les recommandations du DCMI en utilisant autant que possible des types de données standards pour les différents champs de notre modèle (ex. dates et intervalles temporels au format W3CDTF) et surtout en nous appuyant sur des vocabulaires standards ou ontologies du domaines lorsqu'ils existent. Ainsi, nous utilisons (i) pour renseigner les sujets traités par une source de données, le thésaurus biomédical MeSH maintenu par la NLM[2] ; (ii) pour renseigner les organismes couverts par une source, une ontologie extraite de la taxonomie des organismes vivants du NCBI utilisée notamment pour annoter les séquences de GenBank et EMBL. Cette taxonomie étant très riche, nous en avons extrait une ontologie présentée figure 1 en partant des feuilles correspondant aux organismes modèles en biologie (tels la souris, la levure de bière, le riz etc.) jusqu'à la racine (tout organisme) en ne conservant que les nœuds structurants. En supposant que chaque nœud de l'ontologie soit défini par les propriétés communes aux groupes d'organismes correspondants aux nœuds fils, la relation entre les nœuds est une relation de spécialisation (relation d'ordre partiel).

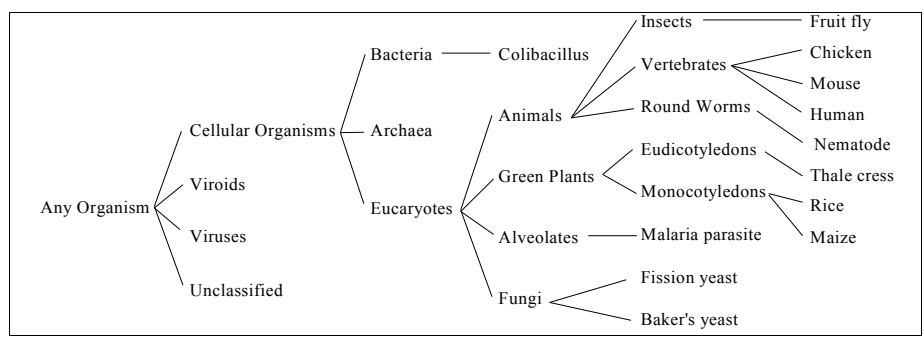

**Figure 1.** *Ontologie des organismes vivants (définie pour BioRegistry)*

---

2. http://www.nlm.nih.gov/mesh/

Une sous-hiérarchie du modèle de BioRegistry est consacrée à la documentation des ontologies utilisées (identifiant, version, date, localisation...). Le modèle de BioRegistry est ouvert et peut intégrer d'autres ontologies du domaine pouvant coexister grâce à un système de préfixation d'un terme d'indexation par la référence de l'ontologie de laquelle il est issu.

Nous avons implémenté le modèle hiérarchique de méta-données de BioRegistry sous la forme d'un schéma XML. Des exemples de documents XML décrivant des sources de données biologiques dans notre modèle peuvent être consultés sur le site web consacré au BioRegistry[3].

### 3.3. *Interrogation de BioRegistry*

Une première forme d'exploitation directe du BioRegistry est une interrogation sur la base d'un formulaire permettant au biologiste d'entrer les valeurs de quelques champs de méta-données et de se voir proposer une liste triée de sources de données du BioRegistry répondant à sa requête. Néanmoins cette approche oblige l'utilisateur à construire une requête, ce qui peut se révéler inefficace sans une vision globale des sources de données contenues dans le BioRegistry.

Afin de dépasser cette limite, nous avons résolu d'appliquer la FCA au contenu du BioRegistry afin de rendre possible (i) une classification flexible des sources de données sur la base du partage de propriétés et (ii) une interrogation du BioRegistry.

Par rapport à l'application de la FCA à un problème de recherche d'information traditionnel, l'avantage ici est que le BioRegistry contient beaucoup moins d'items (de l'ordre d'un millier de sources) que la plupart des corpus documentaires, limitant ainsi l'espace de recherche et la complexité du traitement de la requête. Nous considérons ceci comme une condition suffisante pour un passage à l'échelle.

Les ontologies de domaine utilisées pour certains champs du BioRegistry (voir section précédente) sont utilisables par le biologiste pour sélectionner les termes de sa requête. Elles seront également exploitées pour le raffinement de la requête en cas d'échec afin d'améliorer le rappel (section 5.2).

### 4. Treillis de concepts pour la classification des sources de données de BioRegistry

#### 4.1. *Construction du treillis de concepts de BioRegistry*

Dans cette section nous détaillons l'application de la FCA au contenu de BioRegistry. Des définitions plus détaillées sur l'analyse de concepts formels sont présentées dans (Ganter et al., 1999).

---

3. http://bioinfo.loria.fr/Members/devignes/Bioregistry/presentationBioregistry/view

Dans la suite, la formalisation de BioRegistry est donnée par le contexte formel $\mathcal{K}_{bio} = (G, M, I)$ où $G$ est un ensemble de sources de données biologiques (ex. Swissprot, RefSeq,...), $M$ est un ensemble de méta-données (ex. *manual revision*, *human organism*,...) et $I$ est une relation binaire entre $G$ et $M$ appelée incidence de $\mathcal{K}_{bio}$ et vérifiant : $I \subseteq G \times M$ et $(g, m) \in I$ (ou $gIm$) où $g, m$ sont tels que $g \in G$ et $m \in M$ signifie que la source de données $g$ possède la méta-donnée $m$. Un exemple de contexte formel est donné dans le tableau 1. Les noms complets des sources de données biologiques et ceux des méta-données sont donnés dans le tableau 2 (Les symboles et les abréviations sont utilisés pour une meilleure visibilité). Soit $A \subseteq G$ un ensemble de sources de données. L'ensemble de méta-données communes à toutes les sources dans $A$ est $A' = \{m \in M \mid \forall g \in A, gIm\}$. De façon duale, pour un ensemble de méta-données $B \subseteq M$, l'ensemble des sources possédant toutes les méta-données dans $B$ est $B' = \{g \in G \mid \forall m \in B, gIm\}$. Un concept formel dans la formalisation de BioRegistry $\mathcal{K}_{bio}$ est un ensemble de sources de données partageant un ensemble de méta-données. Il est formellement représenté par un couple $(A, B)$ où $A \subseteq G$, $B \subseteq M$, $A' = B$ et $B' = A$ ; $A$ et $B$ sont respectivement appelés *extension* et *intension* du concept.

| Sources\Méta-données | NS | PS | AO | An | Ve | Hu | Mo | MR |
|---|---|---|---|---|---|---|---|---|
| S1 | 0 | 1 | 1 | 0 | 0 | 0 | 0 | 1 |
| S2 | 1 | 1 | 1 | 0 | 0 | 0 | 0 | 1 |
| S3 | 1 | 0 | 0 | 0 | 0 | 1 | 0 | 0 |
| S4 | 0 | 1 | 1 | 0 | 0 | 0 | 0 | 1 |
| S5 | 1 | 1 | 0 | 0 | 0 | 1 | 0 | 0 |
| S6 | 1 | 0 | 0 | 1 | 0 | 0 | 0 | 0 |
| S7 | 0 | 1 | 0 | 0 | 0 | 0 | 1 | 0 |
| S8 | 0 | 1 | 0 | 0 | 1 | 0 | 0 | 0 |

**Tableau 1.** *Un exemple de contexte formel Bioregistry ($\mathcal{K}_{bio} = (G, M, I)$)*

| Nom de la source | Symbole |
|---|---|
| Swissprot | S1 |
| RefSeq | S2 |
| TIGR-HGI | S3 |
| GPCRDB | S4 |
| HUGE | S5 |
| ENSEMBL | S6 |
| Mouse Genome DB | S7 |
| Vega Genome Browser | S8 |

| Méta-données | Abréviation | Catégorie |
|---|---|---|
| Nucleic sequence | NS | Subject |
| Proteic sequence | PS | Subject |
| Any organism | AO | Organism |
| Animals | An | Organism |
| Vertebrate | Ve | Organism |
| Human | Hu | Organism |
| Mouse | Mo | Organism |
| Manual revision | MR | Quality |

**Tableau 2.** *Noms Complets des sources de données biologiques et de leurs méta-données*

On note par $C$ l'ensemble des concepts de $\mathcal{K}_{bio}$. Soient $C1 = (A1, B1)$ et $C2 = (A2, B2)$ dans $C$, $C1$ est subsumé par $C2$ (noté par $C1 \sqsubseteq C2$) si $A1 \subseteq A2$ ou de façon duale $B2 \subseteq B1$. $(C, \sqsubseteq)$ est un treillis complet (Wille, 1984) appelé treillis de concepts correspondant au contexte formel $\mathcal{K}_{bio}$. On notera dans la suite $(C, \sqsubseteq)$ par $\mathcal{L}(C)$. La figure 2 montre le treillis de concepts $\mathcal{L}(C)$ correspondant au contexte formel de BioRegistry $\mathcal{K}_{bio}$ donné par le tableau 1.

Une caractéristique importante du contexte formel $\mathcal{K}_{bio}$ est que l'ensemble $M$ de méta-données est soigneusement choisi durant la construction de BioRegistry. Ceci permet d'éviter l'explosion de la complexité du contexte. Cette particularité nous motive dans le choix de l'algorithme de Godin (Godin et al., 1995) pour construire le treillis de concepts correspondant puisque le contexte est petit et peu dense (Kuznetsov et al., 2002). De plus, comme déjà mentionné dans la section 2.3, cet algorithme permet l'ajout de nouveaux concepts à un treillis existant. Cet aspect est essentiel pour notre méthode d'interrogation qui sera décrite dans la section 5.

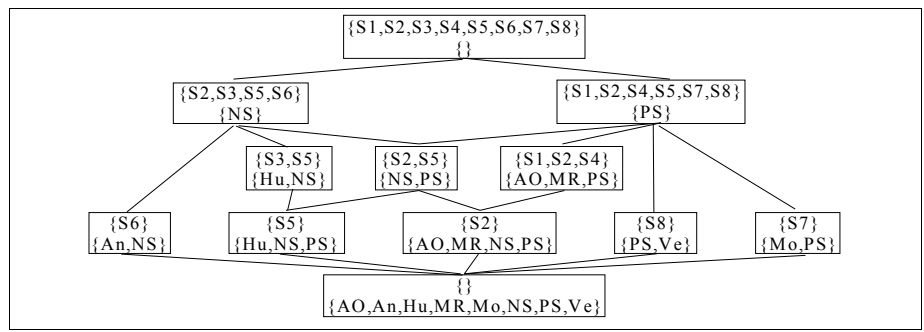

**Figure 2.** *Le treillis de concepts $\mathcal{L}(C)$ correspondant au contexte formel $\mathcal{K}_{bio}$*

### 4.2. *Classification flexible des sources de données de BioRegistry*

Le modèle structuré de BioRegistry permet d'en extraire facilement des ensembles variés de sources de données et/ou de méta-données, qui peuvent conduire grâce à la FCA à une grande diversité de vues sur l'organisation générale des sources de données biologiques. Par exemple un utilisateur intéressé par les méta-données de type *Subjects* (voir section 3.2) associées à l'ensemble des sources pourra définir un contexte formel modifié où les attributs ne seront constitués que des méta-données de la catégorie *Subjects*. L'ensemble d'objets correspondants comprendra toutes les sources de données pour lesquelles au moins une méta-donnée de cette catégorie a été renseignée. Un treillis pourra alors être construit, permettant de visualiser le partage des méta-données de la catégorie *Subjects* dans BioRegistry.

Dans un autre cas, l'utilisateur sera intéressé par visualiser la classification des sources de données relatives à l'organisme humain. Un nouveau contexte formel sera construit automatiquement dans lequel l'ensemble des objets comprend toutes les sources de données de BioRegistry dont la méta-donnée *Organism* a la valeur *Human*. L'ensemble des attributs sera constitué de toutes les méta-données associées à ce jeu de sources. Un nouveau treillis sera alors construit répondant aux besoins de l'utilisateur.

## 5. Interrogation du treillis de concepts de BioRegistry

### 5.1. *Recherche des sources de données biologiques pertinentes*

Une fois que le treillis de concepts $\mathcal{L}(C)$ est construit, la recherche des sources de données pertinentes peut commencer. De la même façon que (Godin et al., 1995) et (Carpineto et al., 2000), nous définissons un concept requête $Q = (Q_A, Q_B)$ où $Q_A = \{Query\}$, càd un nom pour l'extension recherchée (il peut aussi être vu comme un nom pour désigner une extension vide ou une classe virtuelle à instancier) et $Q_B$ est l'ensemble de méta-données utilisées pour la recherche. Considérons comme exemple la requête cherchant les sources de données ayant les méta-données *Nucleic Sequences*, *Human* et *Manual Revision*. En utilisant les abréviations données dans le tableau 2, le concept de la requête est $Q = (\{Query\}, \{NS, Hu, MR\})$.

Une fois défini, le concept $Q$ est inséré dans le treillis $\mathcal{L}(C)$ en utilisant l'algorithme de construction incrémentale de Godin (Godin et al., 1995). Le treillis résultant est noté par $(C \oplus Q, \sqsubseteq)$ où $(C \oplus Q)$ désigne le nouvel ensemble de concepts résultant de l'insertion de la requête. Dans ce qui suit, $(C \oplus Q, \sqsubseteq)$ sera noté par $\mathcal{L}(C \oplus Q)$. Le treillis $\mathcal{L}(C \oplus Q)$ correspondant à l'exemple mentionné précédemment est représenté dans la figure 3. Les concepts représentés par les rectangles en pointillé sont soit de nouveaux concepts soit des concepts modifiés suite à l'insertion de la requête dans le treillis. Ces concepts sont les seuls qui partagent des méta-données avec la requête et peuvent ainsi contenir des sources de données pertinentes pour l'utilisateur.

Il est à noter que, dans cet exemple, le concept requête résultant (noté aussi $Q$) dans $\mathcal{L}(C \oplus Q)$ est identique au concept requête initial $Q$. Dans le cas général, s'il existe dans $\mathcal{L}(C)$ un concept de la forme $(A, Q_B \cup B)$ alors l'insertion de $Q$ dans $\mathcal{L}(C)$ va produire un concept de la forme $(\{Query, A\}, Q_B)$ qui sera considéré comme le nouveau concept requête. Pour plus de simplicité, nous continuons à noter par $Q$ le concept requête $\mathcal{L}(C \oplus Q)$ dans tous les cas.

DÉFINITION. — Une source de donnée est pertinente pour une requête donnée si et seulement si elle partage au moins l'une de ses méta-données avec la requête. Le degré de pertinence d'une source de donnée est donné par le nombre de méta-données qu'elle partage avec la requête et par l'étape durant laquelle la source de données a été ajoutée au résultat.

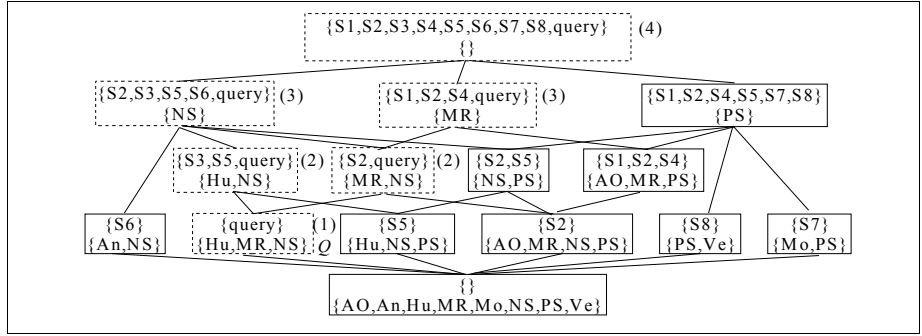

**Figure 3.** *Le treillis de concepts* $\mathcal{L}(C \oplus Q)$

Cette définition de pertinence est la base de notre processus de recherche. Elle est différente de la notion de voisinage utilisée dans (Carpineto et al., 2000). En effet, celle-ci peut aboutir à l'obtention de documents qui ne partagent aucun terme avec la requête, ce qui ne correspond pas à nos besoins.

Considérant la définition précédente, toutes les sources de données pertinentes sont dans l'extension de $Q$ et de ses subsumants dans le treillis de concepts (les concepts représentés en pointillé dans la figure 3) puisque l'intension de chacun de ces concepts est incluse dans $Q_B$ (l'intension du concept requête). Dans ce qui suit nous notons par $\mathcal{R}_{sources}$ l'ensemble des sources de données pertinentes pour la requête considérée. Il est important de mentionner ici que toutes les sources présentes dans $\mathcal{R}_{sources}$ n'ont pas forcément le même degré de pertinence. En effet, elles sont ordonnées selon le nombre de méta-données qu'elles partagent avec la requête et selon l'étape durant laquelle elles ont été ajoutées à $\mathcal{R}_{sources}$.

Intuitivement, l'algorithme de recherche de sources de données pertinentes consiste d'abord à insérer le concept requête dans le treillis $\mathcal{L}(C)$. Ensuite, l'ensemble des sources de données apparaissant dans l'extension de $Q$ dans le treillis $\mathcal{L}(C \oplus Q)$ (si elles existent) sont ajoutées au résultat $\mathcal{R}_{sources}$. Soit $C_1$ l'ensemble des subsumants directs (à distance 1) de $Q$ dans $\mathcal{L}(C \oplus Q)$. L'ensemble de sources de données qui apparaissent dans les extensions des concepts de $C_1$ et qui ne sont pas déjà dans $\mathcal{R}_{sources}$ sont ajoutées au résultat à cette étape. L'étape suivante consiste à déterminer $C_2$ l'ensemble des subsumants des concepts de $C_1$ (ou encore les subsumants de distance 2 de la requête). De la même façon que pour $C_1$, les nouvelles sources dans les extensions des concepts de $C_2$ sont ajoutées au résultats $\mathcal{R}_{sources}$. La même opération est effectuée jusqu'à atteindre un ensemble $C_n$ qui est vide. À chaque étape, les sources de données apparaissant dans un concept à intension vide sont ignorées. Le rang d'une source dans $\mathcal{R}_{sources}$ peut être mémorisé selon la distance de la source (où du premier concept dans lequel apparaît la source) à la requête dans $\mathcal{L}(C \oplus Q)$.

Dans la figure 3, les numéros à côté des concepts représentent les itérations de l'algorithme expliqué précédemment. Dans la première itération le concept requête $Q$ est considéré. Dans cet exemple l'extension de $Q$ est vide. Aucune source n'est ajoutée au résultat à cette étape. La deuxième itération permet d'ajouter au résultat les sources de données *S3*, *S5* et *S2*. À la troisième itération les sources *S1*, *S4* et *S6* sont ajoutées à $\mathcal{R}_{sources}$. À la quatrième itération l'intension du concept est vide. Aucune source n'est donc ajoutée à cette étape et l'algorithme s'arrête. Ainsi $\mathcal{R}_{sources}$ est constitué des sources *S3* et *S5* (grâce à *Hu* et *NS*), *S2* (grâce à *NS* et *MR*) au premier rang, et *S6* (grâce à *NS*) et *S1* et *S4* (grâce à *MR*) au deuxième rang.

Des critères supplémentaires relatifs à des préférences données peuvent être considérés pour raffiner l'ordre des sources de données dans le résultat $\mathcal{R}_{sources}$.

**5.2. *Raffinement de requête à partir d'ontologies de domaine***

La réponse à une requête donnée peut être vide. Par exemple dans le contexte formel de BioRegistry donné dans le tableau 1, une requête cherchant une source de données contenant des données relatives à l'organisme *Chicken* aura un résultat $\mathcal{R}_{sources}$ nul. Cependant, des sources de données pertinentes à cette requête peuvent exister mais les méta-données qui les décrivent ne correspondent pas exactement à ce qui est exprimé dans la requête. Pour remédier à cette limite nous proposons le mécanisme de raffinement de requête à partir d'ontologies de domaines.

Contrairement aux propositions (Carpineto et al., 2000, Priss, 2000, Safar et al., 2004) mentionnées dans la section 2.2, nous modifions la requête au lieu du treillis. En effet, nous préservons la structure entière du treillis et nous modifions la requête en y ajoutant des méta-données en relation sémantiques avec les méta-données de la requête initiale à partir d'une ontologie donnée. Cette stratégie automatisable évite l'introduction de la redondance dans le treillis.

Les méta-données ajoutées à la requête sont soit plus spécifiques soit plus générales que celles de la requête initiale. Ceci nous amène à définir deux types de raffinement : raffinement par généralisation et raffinement par spécialisation. Il est important de rappeler ici que nous ne sommes pas confrontés à des problèmes de synonymie entre les méta-données de la requête et les termes de l'ontologie puisque les méta-données utilisées lors de la construction de BioRegistry sont des termes extraits d'ontologies du domaine.

Le raffinement par généralisation relatif à l'une des méta-données de la requête permet d'ajouter les méta-données plus générales représentées par les ancêtres de la méta-donnée considérée dans l'ontologie. Dans l'exemple cité précédemment (méta-donnée *Chicken*) et en considérant l'ontologie donnée dans la figure 1, les méta-données qui peuvent être ajoutées à la requête lors du raffinement par généralisation sont *Vertebrates, Animals, Eucaryotes, Cellular Organisms*, et *Any Organism*. Cependant, quelques une parmi ces méta-données ne figurent pas dans le contexte formel $\mathcal{K}_{bio}$ donné par le tableau 1 (*Eucaryotes* et *Cellular Organisms*). Ceci signifie que ces méta-données ne sont partagées par aucune source de données dans ce contexte ce qui rend inutile leur ajout à la requête puisqu'il n'enrichit pas le résultat. Seules les méta-données présentes dans $\mathcal{K}_{bio}$ seront considérées durant le processus de raffinement par généralisation.

De façon duale le raffinement par spécialisation relativement à une méta-donnée de la requête permet l'ajout des méta-données plus spécifiques représentées par les descendants de la méta-donnée considérée dans l'ontologie. Dans l'exemple précédant la méta-donnée *Chicken* n'a pas de descendant dans l'ontologie ce qui rend impossible le raffinement de la requête pour cette méta-donnée. Un meilleur exemple pour illustrer ce type de raffinement sera de considérer la requête composée de la méta-donnée *Eucaryotes* qui ne retrouve aucune réponse puisqu'elle n'apparaît pas dans le contexte formel $\mathcal{K}_{bio}$. Le raffinement par spécialisation permet d'ajouter à la requête tous les descendants de la méta-donnée *Eucaryotes* dans l'ontologie, qui apparaissent dans le contexte formel (*Animals*, *Vertebrate*, *Human* et *Mouse*).

Il est possible de combiner les deux types de raffinement de requête en ajoutant à la requête à la fois les descendants et les ancêtres de la méta-donnée considérée dans l'ontologie. Dans tous les types de raffinement, le nombre de méta-données ajoutées à la requête peut être contrôlé en considérant uniquement les ancêtres les plus proches de la méta-donnée considérée dans l'ontologie (dans le cas du raffinement par généralisation) et/ou ses descendants les plus proches (dans le cas du raffinement par spécialisation).

Une fois le raffinement de requête effectué, la requête raffinée est insérée dans le treillis initial $\mathcal{L}$ ($C$) et l'algorithme décrit précédemment est appliqué au treillis résultant $\mathcal{L}$ ($C \oplus Q$). Les sections qui suivent présentent l'apport du mécanisme de raffinement de requête à partir d'ontologies de domaines.

### 5.3. *Raffinement de requête par généralisation*

Considérons la requête constituée de la méta-donnée *Chicken* et représentée par le concept formel $Q = (\{Query\}, \{Ch\})$. La réponse à cette requête est vide puisque cette méta-donnée n'apparaît pas dans le contexte formel. Après le recours au raffinement par généralisation de cette requête, nous obtenons comme réponse à la requête raffinée les sources *S6* (grâce à *Animals*), *S8* (grâce à *Vertebrate*) et *S1*, *S2* et *S4* (grâce à *Any Organism*) toutes au premier rang.

Chaque source de données du résultat précédent comprend une partie qui correspond à ce qui est demandé dans la requête (ex. *S8* concernée par *Chicken* mais aussi par *Mouse* et *Human*). Plus la distance entre la méta-donnée de la requête initiale et celle ajoutée lors du raffinement est courte, plus la source résultante est pertinente (*S8* est préférée à *S6*). Cela justifie l'intérêt de contrôler les méta-données ajoutées lors du raffinement par généralisation de requêtes. Ainsi, pour éviter l'introduction des sources de faible pertinence dans le résultat, on ne considère que les ancêtres les plus proches de la méta-donnée considérée.

### 5.4. *Raffinement de requête par spécialisation*

Considérons la requête constituée de la méta-donnée *"Eucaryotes"* et représentée par le concept formel $Q = (\{Query\}, \{Eu\})$. La réponse à cette requête est vide puisque cette méta-donnée n'apparaît pas dans le contexte formel. Après le recours au raffinement par spécialisation de cette requête, nous obtenons comme réponse à la requête raffinée les sources *S6* (grâce à *Animals*), *S8* (grâce à *Vertebrate*), *S5* (grâce à *Human*) et *S7* (grâce à *Mouse*) toutes au premier rang.

Dans ce cas chaque source de données apparaissant dans le résultat donne une réponse partielle à la requête. La composition de ces sources de données permet d'obtenir une réponse complète si chaque descendant de la méta-donnée de la requête indexe une source de données. De façon similaire au raffinement par généralisation, la distance entre la méta-donnée considérée dans la requête initiale et ses descendants dans l'ontologie reflète le degré de pertinence des sources de données présentes dans le résultat. En effet, les sources de données relatives aux méta-données très distantes de la méta-donnée de la requête donnent des informations très précises qui ne sont pas toujours pertinentes pour l'utilisateur. Le degré de spécialisation peut être contrôlé en ne considérant que les descendants dans l'ontologie les plus proches de la méta-donnée de la requête et constituant la meilleure couverture de la requête.

### 5.5. *Choix entre raffinement par généralisation et par spécialisation*

Dans le cas où la méta-donnée considérée est une feuille dans l'ontologie ou sa racine, le problème de choix ne se pose pas puisque dans les deux cas un seul type de raffinement est possible (raffinement par généralisation pour le premier cas et par spécialisation pour le second cas). Cependant, lorsque la méta-donnée considérée n'est ni une feuille ni la racine de l'ontologie, les deux types de raffinement sont possibles et le choix peut être fait selon les préférences de l'utilisateur. En effet, si l'utilisateur accepte des sources de données correspondant en partie à sa requête alors le raffinement par généralisation est adopté et s'il accepte des sources de données correspondant à une partie de sa requête le raffinement par spécialisation est adopté. Dans les deux cas, il est utile de procéder à un ordonnancement a posteriori des sources de données nouvellement ajoutées au résultat (suite au raffinement de la requête). Cet ordonnancement doit être basé sur la similarité entre les méta-données indexant ces sources, d'une part et les méta-données constituant la requête, d'autre part (Ganesan et al., 2003).

## 6. Conclusion et perspectives

Dans ce papier, nous avons présenté une approche combinant FCA et ontologies de domaines pour résoudre un problème de recherche d'information particulier en bioinformatique. BioRegistry, en tant qu'annuaire structuré de méta-données relatives aux sources de données biologiques, constitue un domaine d'application approprié pour la théorie de la FCA et ses potentialités en termes de passage à l'échelle et de flexibilité. L'approche décrite vise à rechercher pour les interroger, des sources de données biologiques pertinentes pour une question donnée. La construction de treillis de concepts apparaît comme un moyen de fournir des vues personnalisées sur les sources de données biologiques et d'organiser les connaissances sur ces sources. De plus un mécanisme de raffinement de requête basé sur les ontologies est proposé pour améliorer le processus de recherche d'information.

Remerciements

**7. Bibliographie**

Carpineto C., Romano G., « A Lattice Conceptual Clustering System and Its Application to Browsing Retrieval. », *Machine Learning*, vol. 24, n° 2, 1996, p. 95-122.

Carpineto C., Romano G., « Order-theoretical ranking », *Journal of the American Society for Information Science*, vol. 51, n° 7, 2000, p. 587-601.

Carmel D., Farchi E., Petruschka Y., Soffer A., « Automatic query refinement using lexical affinities with maximal information gain », *SIGIR'02 : Proceedings of the 25th annual international ACM SIGIR conference on Research and development in information retrieval*, Tampere, Finland, 2002, ACM Press, p. 283-290.

Davidson S., Crabtree J., Brunk B., Brian P., Schug J., Tannen V., Overton G. C., Stoeckert C. J., « K2/Kleisli and GUS : experiments in integrated access to genomic data sources », *IBM systems journal*, vol. 40, 2000, p. 512-531.

Dekkers M., Weibel S., « State of the Dublin Core Metadata Initiative », *D-Lib Magazine*, vol. 9, n° 4, 2003.

Discala C., Benigni X., Barillot E., Vaysseix G., « DBCAT : a catalog of 500 biological databases », *Nucleic Acids Research*, vol. 28, n° 1, 2000, p. 8-9.

Galperin M. Y., « The Molecular Biology Database Collection : 2005 update », *Nucleic Acids Research*, vol. 33, 2005, National Center for Biotechnology Information and National Library of Medicine and National Institutes of Health.

Ganter B., Wille R., *Formal Concept Analysis*, Springer, mathematical foundations édition, 1999.

Ganesan P., Garcia-Malia H., Widom J., « Exploiting hierarchical domain structure to compute similarity », *ACM Transactions on Information Systems*, vol. 21, n° 1, 2003, p. 64-93.

Goble C. A., Stevens R., NG G., Bechhofer S., Paton N. W., Baker P. G., Peim M., Brass A., « Transparent Access to Multiple Bioinformatics Information Sources », *IBM Systems Journal Special issue on deep computing for the life sciences*, vol. 40, n° 2, 2001, p. 532-552.

Godin R., Mineau G., Missaoui R., « Méthodes de classification conceptuelle basées sur les treillis de Galois et applications », *Revue d'intelligence artificielle*, vol. 9, n° 2, 1995, p. 105-137.

Godin R., Missaoui R., Alaoui H., « Incremental Concept Formation Algorithms Based on Galois (Concept) Lattices. », *Computational Intelligence*, vol. 11, 1995, p. 246-267.

Köhler J., Philippi S., Lange M., « SEMEDA : ontology based semantic integration of biological databases », *Bioinformatics*, vol. 19, 2003, p. 2420-2427.

Kuznetsov S., Obiedkov S., « Comparing Performance of Algorithms for Generating Concept Lattices », *Journal of Experimental & Theoretical Artificial Intelligence*, vol. 14, 2002, p. 189-216, Taylor & Francis.

Lord P., Bechhofer S., Wilkinson M. D., Schiltz G., Gessler D., Hull D., Goble C., Stein L., « Applying semantic web services to Bioinformatics : Experiences gained, lessons lernt », *ISWC*, 2004.

Oinn T., Addis M., Ferris J., Marvin D., Senger M., Greenwood M., Carver T., Glover K., Pocock M. R., Wipat A., LI P., « Taverna : a tool for the composition and enactment of bioinformatics workflows », *Bioinformatics*, vol. 20, 2004, p. 3045-3054.

Pernelle N., Rousset M.-C., Soldano H., Ventos V., « ZooM : a nested Galois lattices-based system for conceptual clustering », *Journal of Experimental and Theoritical Artifial Intelligence (JETAI)*, vol. 14, n° 2, 2002, p. 157-187.

Priss U., « Lattice-based Information Retrieval », *Knowledge Organization*, vol. 27, n° 3, 2000, p. 132-142.

Safar B., Kefi H., Reynaud C., « OntoRefiner : a user query refinement interface usable for Semantic Web Portals », *Proceedings of the ECAI 2004, Workshop on Application of Semantic Web Technologies to Web Communities*, August 2004.

Stumme G., Taouil R., Bastide Y., Lakhal L., « Conceptual Clustering with Iceberg Concept Lattices », *Proceeding GI-Fachgruppentreffen Maschinelles Lernen (FGML'01)*, Oktober 2001.

Wille R., « Restructuring lattice theory : an approach based on hierarchies of concepts », *Orderd sets*, vol. 23, 1982, p. 445-470, Rival editor.

Wille R., « Line diagrams of hierarchical concept systems », *International Classification*, vol. 2, 1984, p. 77-86.

Wroe C., Stevens R., Goble C., Roberts A., Greenwood M., « A suite of DAML+OIL Ontologies to Describe BioinformaticsWeb Services and Data », *International Journal of Cooperative Information Systems special issue on Bioinformatics*, 2003.